# Analysis of beta-decay data acquired at the Physikalisch-Technische Bundesanstalt: evidence of a solar influence

P.A. Sturrock[a], G. Steinitz[b], E. Fischbach[c], A. Parkhomov[d], J.D. Scargle[e]
[a] Kavli Institute for Particle Astrophysics and Cosmology and Center for Space Science and Astrophysics, Stanford University, Stanford, CA 94305-4060, USA
[b] Geological Survey of Israel, Jerusalem, 95501, Israel
[c] Department of Physics and Astronomy, Purdue University, West Lafayette, IN 47907, USA
[d] Institute for Time Nature Explorations, Lomonosov Moscow State University, Moscow, Russia
[e] NASA/Ames Research Center, MS 245-3, Moffett Field, CA 94035, USA

*Corresponding author. Tel +1 6507231438; fax +1 6507234840.
Email address: sturrock@stanford.edu

ABSTRACT
According to an article entitled *Disproof of solar influence on the decay rates of 90Sr/90Y* by Kossert and Nähle of the Physikalisch-Technische Bundesanstalt (PTB) [1], the PTB measurements show no evidence of variability. We show that, on the contrary, those measurements reveal strong evidence of variability, including an oscillation at 11 year$^{-1}$ that is suggestive of an influence of internal solar rotation. An analysis of radon beta-decay data acquired at the Geological Survey of Israel (GSI) Laboratory for the same time interval yields strong confirmation of this oscillation.

# 1. Introduction

In recent years, a number of articles have been published presenting evidence that some beta-decay rates are variable. Falkenberg, writing in 2001, reported evidence of an annual variation in the decay rate of tritium and suggested an association with the varying Earth-Sun distance [2]. This article was criticized by Bruhn [3], to which Falkenberg responded in a further article [4]. Such interchanges have recurred not infrequently. Jenkins and Fischbach [5,6] reviewed the experimental results of Alburger et al. of the Brookhaven National Laboratory [7] concerning the decay rates of $^{32}$Si and $^{36}$Cl, and of data acquired at the Physikalisch-Technische Bundesanstalt (PTB) in Braunschweig, Germany, concerning the decay rate of $^{226}$Ra [8]. Like Falkenberg, Jenkins and Fischbach proposed a relationship to the varying Earth-Sun distance. The Jenkins-Fischbach articles led to critical articles by Cooper [9], Norman [10] and Semkow [11], which led to responses by Krause et al. [12], O'Keefe et al. [13], and Jenkins et al. [14].

The variability of beta-decay rates has more recently been called into question by Kossert and Nahle (KN) of PTB [1]. KN base their concerns on their power-spectrum analysis of measurements of the decay of $^{90}$Sr/$^{90}$Y using the TDCR (Triple-to-Double Coincidence Ratio) experimental method [15 - 17]. Their results appear to contradict the positive results of earlier experiments by one of us (AP) [18].

KN have kindly made their measurements available to us for independent analysis. In Section 2, we carry out a power-spectrum analysis of the PTB measurements and assess the statistical significance of the principal peaks in the resulting power spectra. In Section 3, we discuss the difference between the KN significance estimates and our estimates. In Section 4, we carry out a power-spectrum analysis of beta-decay data, for the same time interval, extracted from data compiled at the Geological Survey of Israel Laboratory [19-21]. We carry out spectrogram analyses in Section 5, we discuss our results in Section 6, and we summarize our conclusions in Section 7. We present some of the basic information about the PTB experiment in the Appendix.

# 2. Power spectrum analysis

KN have investigated possible variations in the decay of $^{90}$Sr/$^{90}$Y sources by using the TDCR method that has been developed by standards laboratories as a way to arrive at accurate estimates of absolute decay rates [15 - 17]. This experiment therefore differs significantly from all other experiments designed to study the possible variability of beta decays. All other experiments simply measure the count rates of nuclides. The PTB experiment measured the *triple coincidences* of decay events as registered by three photo-multiplier tubes (PMTs). We show the layout of the PMTs and some of the characteristics of the measurement procedure in the Appendix.

KN made sequential measurements of three samples (S2, S3 and S4) and also of a blank sample (S1) to monitor environmental effects. KN derived "activity" estimates, shown in KN Figures 4, 5 and 6, from their triple-coincidence measurements and theoretically calculated counting efficiencies.

We show in Figures 1, 2 and 3 power spectra formed from the activity measures by a likelihood procedure [22] which, for present purposes, is equivalent to the Lomb-Scargle procedure [23, 24].

We see that the curves in these figures are very close to those shown in KN Figures 6, 7 and 8. However, the significance estimates are completely different. For example, the biggest peak in Figure 1 is found at frequency 11.32 year$^{-1}$ and has power $S = 8.42$. According to Scargle theory [24], the probability of finding that power or more at a specified frequency is given by

$$P = \exp(-S). \qquad (1)$$

This probability is found to be $2 \times 10^{-4}$.

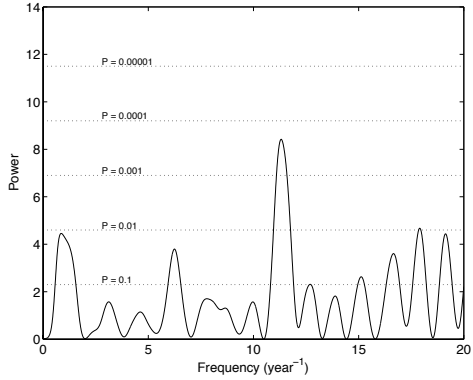

Figure 1. Power spectrum for PTB Sample 2. (c.f. KN Figure 6.) Counting only peaks with powers of 5 or more, we find one peak with power 8.4

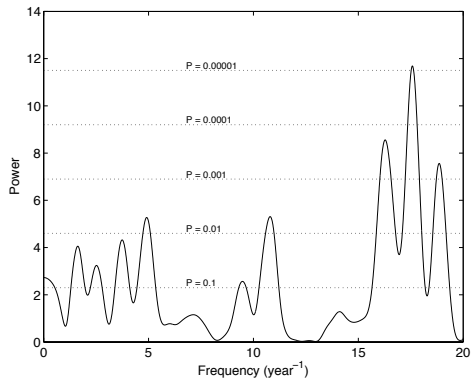

Figure 2. Power spectrum for PTB Sample 3. (c.f. KN Figure 7.) Counting only peaks with powers of 5 or more, we find 5 peaks with powers 11.7, 8.6, 7.6, and two at 5.3.

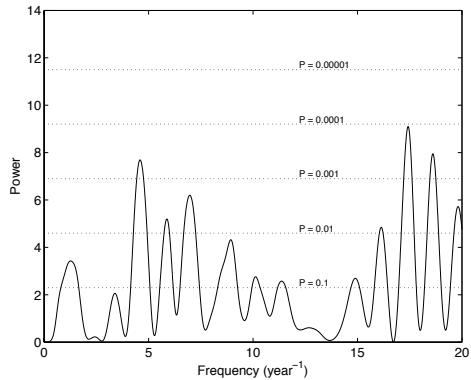

Figure 3. Power spectrum for PTB Sample 4. (c.f. KN Figure 8.) Counting only peaks with powers of 5 or more, we find 6 peaks with powers 9.1, 8.0, 7.7, 6.2, 5.7 and 5.2.

By contrast, KN indicate significance levels by a quantity $\alpha$ which is not defined but appears to be related to the power by

$$\alpha = \exp\left(-S^{1/2}\right). \qquad (2)$$

According to KN Figure 6, the biggest peak (at about 11.2 year⁻¹, with S close to 8.4) is near to the $\alpha = 0.5$ level, leading KN to conclude that the modulation at that frequency is not significant. The power spectra for Samples 3 and 4, shown in Figures 2 and 3 (virtually identical to KN Figures 7 and 8), show even stronger peaks with powers up to 11.7, which corresponds to a statistical significance level of 8 10⁻⁶. However, based on their estimates of the quantity $\alpha$, KN regard all of these peaks as insignificant.

### 3. Comparison of significance estimates

To further check this discrepancy, we have carried out a Monte Carlo calculation, using the shuffle test [25]. Figure 4 shows the reverse-cumulative distribution of the maximum power at a specified frequency (taken to be that of the principal peak in Figure 4, although the choice is not significant) computed from 10,000 shuffles of the data. We find that a fraction 2 10⁻⁴ of the shuffles give powers larger than the actual power (8.42), which is what one would expect from Equation (1), confirming that Equation (1) is indeed the appropriate formula for statistical significance estimation.

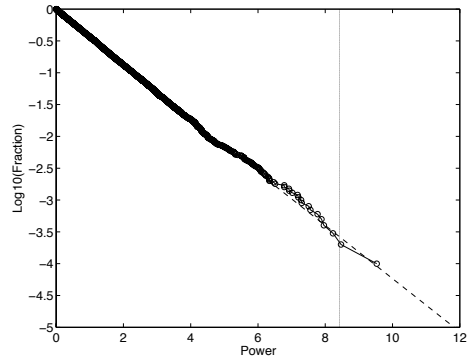

Figure 4. Shuffle test of the principal peak in the power spectrum for PTB Sample 2.

We now make an approximate estimate of the probability of obtaining the power spectra shown in Figure 1, 2 and 3. Counting only peaks with power 5 or more, we find that, for Sample 2, Figure 1 shows 14 peaks of which one has power 8.4. The probability that the strongest peak has power 8.4 is $P_2 = 14 \times e^{-8.4}$, i.e. 0.003. The power spectrum for Sample 3, shown in Figure 2, has 13 peaks, of which 5 have powers of 5 or more: 11.7, 8.6, 7.6, and two at 5.3. The probability that these peaks have occurred by chance may be calculated by multiplying the chain $13 \times e^{-11.7}$, $12 \times e^{-8.6}$, etc., from which we find that $P_3 = 3 \times 10^{-14}$. The power spectrum for Sample 4 shown in Figure 3 has 15 peaks, of which 6 have powers of 5 or more: 9.1, 8.0, 7.7, 6.2, 5.7 and 5.2. A similar calculation leads to the estimate $P_4 = 2 \times 10^{-12}$. Combining these figures, we estimate that the probability of finding by chance the peaks with power 5 or more in Figures 1, 2 and 3 (which are indistinguishable from KN Figures 6, 7 and 8) is of order $10^{-29}$.

As an independent check on the significance of the fluctuations in the PTB data, we have grouped all of the measurements into 50 bins (with equal occupancy). We show the normalized measurements, together with the standard errors of the mean, in Figure 5. This figure confirms that there are significant departures from constancy in the measurements. This analysis lends itself to a chi-square analysis. We find the chi-square value to be 227. The probability of such a large value, for 49 degrees of freedom, is found to be 10⁻²³. This provides further evidence that the PTB measurements have their origin in one or more non-random processes.

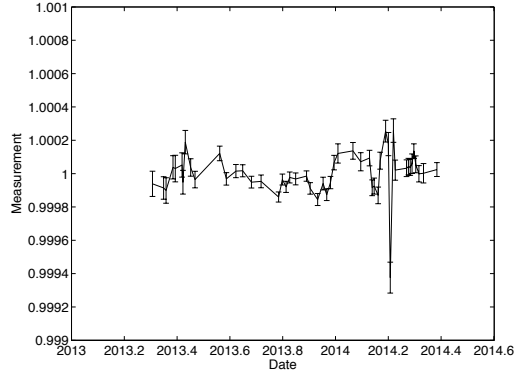

Figure 5. Normalized measurements, with standard error of the mean, of all samples grouped into 50 bins.

Although each of the power spectra shown in Figures 1, 2 and 3 shows statistically significant peaks, the power spectra are all different. This indicates that the power spectra are not responding to the same influence, which would be the case if they were responding only to solar rotational modulation, for instance. This suggests that the experiment is responding to some kind of experimental artifacts and/or some combination of periodic influences.

We would have more confidence in the power spectra if we had error estimates for each measurement, but this information is not available. However, as we see from Figure 5, it is possible to assign a mean value and a standard error of the mean to data organized in bins. This procedure was helpful in visualizing the data and in carrying out a statistical significance estimate. This suggests that it may be interesting to carry out a power spectrum analysis of binned data.

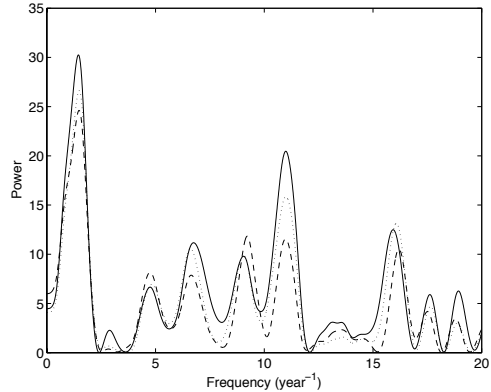

Figure 6. Power spectrum formed from the PTB measurements by a likelihood procedure from all data organized successively into 100 bins (dashed line), 200 bins (dotted line), and 300 bins (solid line).

We have carried out a sequence of calculations, dividing the data successively into 100, 200, etc., bins. For each bin, we calculate the mean, the standard deviation, and the standard error of the mean. For each choice, we carry out a power spectrum analysis, using the likelihood procedure and entering the mean and standard error of the mean of each bin. Figure 6 shows the results for 100, 200, and 300 bins. We see that for some frequencies there is a steady increase in power whereas for other frequencies there is no steady trend. We find the most pronounced trend for the peak at 10.95 year$^{-1}$, for which the power increases by 80% (to S = 20). An oscillation at or near 11 year$^{-1}$ is suggestive of a solar influence, as we discuss in Sections 4 and 5. We therefore show in Figure 7 a reconstruction of the measurements (for 300 bins) for that frequency for a short time interval at the beginning of 2014. We see that the amplitude is 4 10$^{-5}$, and the modulation is a maximum at 2014.01.

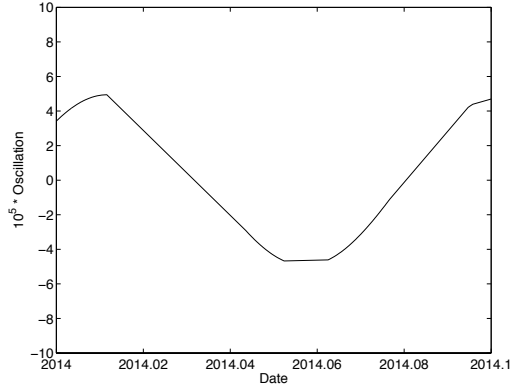

Figure 7. Reconstruction of measurements for 300 bins at frequency 11 year$^{-1}$. The amplitude is 4 10$^{-5}$, and the maximum is at 2014.01.

### 4. Analysis of GSI data

The fact that Figures 6 and 7 show evidence of an oscillation with frequency 11 year$^{-1}$ is suggestive of a solar influence, since oscillations with this frequency have been found in $^{36}$Cl beta-decay data and in Super-Kamiokande neutrino measurements [26, 27]. It is therefore interesting to see if there is independent evidence for an oscillation at this frequency for the same time interval in another dataset.

One of us (GS) has been running a beta-decay experiment without interruption since 28 January 2007 at a laboratory of the Geological Survey of Israel (GSI) in Jerusalem. The experiment measures the beta-decay rate of radon arising from a $^{238}$U source in a closed container. Details of the experiment and progress reports are to be found in refs. [19 - 21]. The following measurements are recorded every hour: date, time, ambient temperature, electrical supply voltage, 3 gamma-detector readings, and 2 alpha-detector readings. The record to date runs to almost 80,000 lines. (GSI measurements relevant to this article are available at http://wso.stanford.edu/pas.)

We have extracted measurements of the central gamma detector for the time interval of the PTB experiment (estimated to be April 21, 2013 to May 31, 2014). The count rate for 2 hours centered on noon is 1,030 per hour and the normalized measurements are shown in Figure 8. The count rate for 2 hours centered on midnight is 960 per hour and the normalized measurements are shown in Figure 9. We see that the noon measurements are more variable than the midnight measurements. The depth of modulation of the count rate is about 6% by day and about 3% by night.

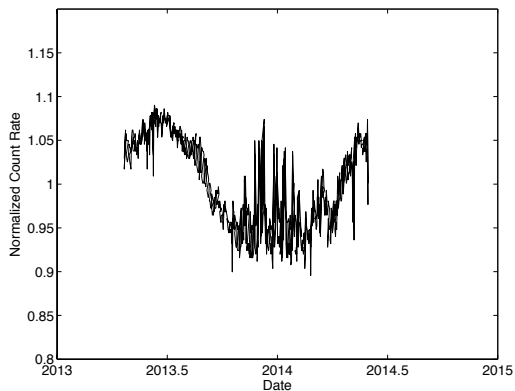

Figure 8. Normalized count rate of the GSI experiment for 2 hours centered on noon, for the time interval of the PTB experiment.

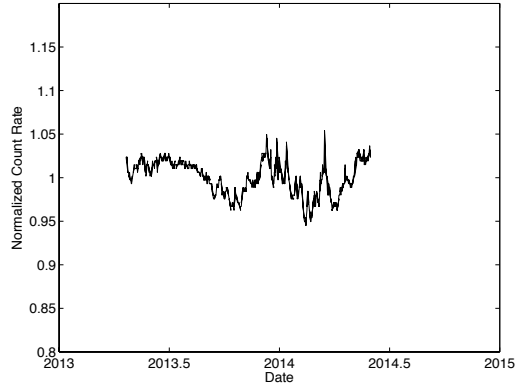

Figure 9. Normalized count rate of the GSI experiment for 2 hours centered on midnight, for the time interval of the PTB experiment.

We show in Figures 10 and 11 power spectra formed from these two time series. There is no evidence of modulations other than the annual modulation in the daytime power spectrum (Figure 10). However, Figure 11 shows evidence of a modulation at 10.95 year$^{-1}$ with power S = 20, the same frequency (that has solar significance) as the prominent PTB oscillation in Figure 6.

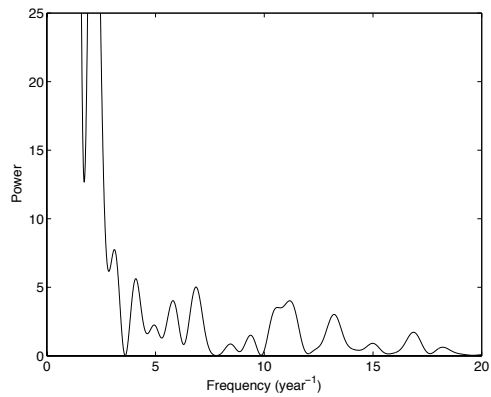

Figure 10. Power spectrum formed from the noon GSI measurements. (The annual oscillation is off chart, with a power S = 290.)

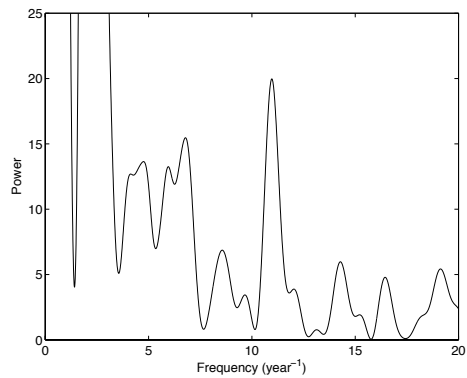

Figure 11. Power spectrum formed from the midnight GSI measurements. (The annual oscillation is off chart with a power S = 100, and there is a semiannual oscillation with power S = 170.)

For comparison with Figure 7, we show in Figure 12 the modulation of the GSI measurements for frequency 11 year$^{-1}$, for the beginning of year 2014. The amplitude is found to be 0.6%, much larger than the amplitude (4 10$^{-5}$) of the modulation of the PTB measurements shown in Figure 7. However, the maximum occurs at 2014.02, quite close to the time of maximum in Figure 7 (2014.01).

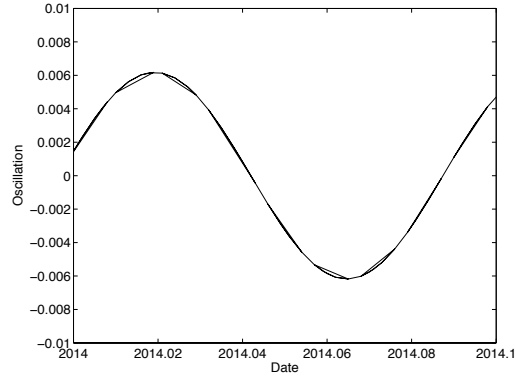

Figure 12. Reconstruction of GSI measurements at frequency 11 year$^{-1}$. The amplitude is 0.006 and the maximum is at 2014.02.

**5. Spectrogram analysis.**
Power-spectrum analysis is well suited to the analysis of systems that are in a steady state or close to a steady state, but this is usually not the case for experiments that investigate beta-decay variability. In this situation, it is advantageous to also use spectrograms that display power spectra as functions of time. (See, for instance, [26,27].) However, the PTB dataset is too short (a little over one year) for this procedure to be feasible.

As a variant of this procedure, we show in Figure 13 a modified spectrogram formed from the PTB dataset, which displays the power of oscillations as a function of frequency and hour of day, as previously used in our study of GSI data [28]. The ordinate is the local time at Braunschweig. This diagram has a complex pattern, but there seem to be two time-of-day intervals with notable oscillations: one from 7 – 10 hours, and the other from 14 – 17 hours.

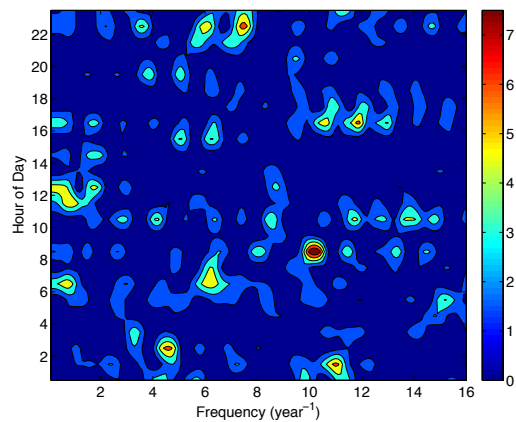

Figure 13. Spectrogram formed from the PTB dataset showing the power of the modulation of the count rate as a function of frequency and hour of day**.**

Since these bands are located symmetrically with respect to noon, this suggests that the measurements may be influenced by the elevation of the Sun with respect to the horizon, as we found to be the case for GSI data [28]. We show in Figure 14 a modified spectrogram in which the ordinate is the solar elevation, which ranges from -55 degrees at midnight in midsummer to +55 degrees at noon in midsummer. We see that Figure 14 has a simpler structure than Figure 13, suggesting that measurements are indeed influenced by the solar elevation. There seem to be two bands in the figure, one from – 30 degrees to -15 degrees, and the other from 12 degrees to 30 degrees.

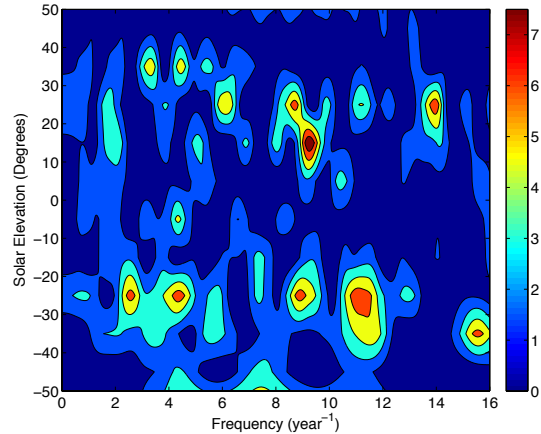

Figure 14. Spectrogram from the PTB dataset showing the power of the modulation of the count rate as a function of frequency and solar elevation.

## 6. Discussion.

Analyses presented in Section 2 yield overwhelming evidence that the TDCR measurements cannot be attributed to a stationary random process, implying that they are subject to one or more unknown complicating factors. One cannot draw firm conclusions about the significance of the TDCR measurements until these factors are understood and if possible eliminated.

However, the analysis that leads to Figure 6 suggests that the TDCR measurements are subject in part to a solar influence that originates in a layer where the synodic rotation frequency (the frequency as seen from Earth) is 11 year$^{-1}$. Analysis of GSI nighttime measurements of the beta decay of $^{222}$Rn leads to the power spectrum shown in Figure 11, which independently shows strong evidence of the same oscillation. Figures 7 and 12 show that both these oscillations have essentially the same phase. However, the amplitude of this oscillation is much smaller for PTB data (4 10$^{-5}$) than for GSI data (6 10$^{-3}$).

As we point out in recent articles [26,27], previous studies of beta-decay data have given evidence of two forms of solar rotational modulation: one in a frequency band near 11 year$^{-1}$, corresponding to a sidereal frequency (as seen from space) of 12 year$^{-1}$, that we attribute to rotation in an inner tachocline, and the other in a frequency band near 12.5 year$^{-1}$ (a sidereal rotation frequency of 13.5 year$^{-1}$) that we attribute to rotation of the radiative zone. We have found independent evidence for the influence of the inner tacholine in r-mode oscillations evident in solar-diameter measurements made from both ground-based and space-based observatories [29, 30]. It therefore appears that the beta-decay process may be influenced by processes in the deep solar interior.

We found that the PTB modulation appears to be localized in two bands of solar elevation, one from – 30 to – 20 degrees, and the other from 20 to 30 degrees. In a recent article [26], we comment on the day-night asymmetry in measurements of the beta decay of $^{222}$Rn acquired at the GSI Laboratory, and suggest that the asymmetry may be attributed to a directionality of the decay process. We specifically suggest that beta-decays may be stimulated by neutrinos and that the outgoing decay products may tend to travel in the direction of travel of the incoming neutrino. In the GSI experiment, the principal detector is located above the radioactive source, so that it responds preferentially to neutrinos that are traveling vertically upward. For this reason, the experiment responds preferentially to solar neutrinos at midnight (the solar neutrinos having traveled through the Earth). We now look into the possibility that the same tendency for directionality may explain the response of the TDCR system indicated in Figure 14.

We suppose that the experiment is responding to beta decays that are stimulated by solar neutrinos, and that the decay products tend to be confined to a cone (of unknown half-angle) centered on the direction of travel of an incoming neutrino. We recall that the TDCR system is set to register only triple-coincidence events. If a neutrino is traveling in the plane of symmetry of the TDCR system, it cannot be traveling towards all three photomultiplier tubes (PM tubes) at the same time, so it is unlikely to lead to a triple-coincidence event. If the neutrino is traveling nearly vertically upwards or downwards, it is unlikely to inject photons into all three PM tubes at the same time, so it is again unlikely to lead to a triple-coincidence event. On the other hand, a neutrino that is traveling at an angle (but not orthogonally) with respect to the plane of the PM tubes may be more likely to lead to a conical array of photons that may be detected by all three PM tubes, and therefore lead to a triple-coincidence event. Hence the pattern found in Figure 14 may be compatible with the hypothesis that some beta decays may be stimulated by neutrinos. Since there is as yet no theory that might explain neutrino-stimulated beta decays, these suggestions are necessarily speculative.

We now comment briefly on the KN article [1]. This article, entitled *Disproof of solar influence on the decay rates of $^{90}Sr/^{90}Y$*, focuses on evidence for an annual oscillation, specifically considering a modulation due to the varying Earth-Sun distance. Since an annual oscillation could be due to environmental factors, we consider it advisable to search instead for evidence of solar rotation. Our analysis of the PTB data appears to yield evidence for such an influence. The amplitude of the solar oscillation at 11 year$^{-1}$ indicated in Figure 7 is approximately $4 \cdot 10^{-5}$, significantly smaller than the amplitude ($6 \cdot 10^{-3}$) of the 11 year$^{-1}$ oscillation that we find in GSI data, shown in Figure 12. This difference may be due in part to the fact that triple-coincidence events are likely to represent a small fraction of all decay events that occur in the experiment.

GSI data are now available for 9 years (beginning 2007.1247), comprising over 80,000 lines. We plan to present an analysis of the entire dataset in the near future.

**7. Conclusions**

The KN article was entitled *Disproof of solar influence on the decay rates of 90Sr/90Y*. We see from our analysis that this proved to be a misinterpretation of their data that arose from use of an inappropriate formula for estimating the significance of peaks in power spectra. When we use the standard formula for significance estimation, we find evidence of strong fluctuations, of uncertain origin, in their data. Their data also show evidence of an oscillation at a frequency attributable to solar rotation. For confirmation of this fact, we show that exactly the same oscillation is present in GSI data acquired during the same time interval as the PTB data. We conclude that although the TDCR procedure is appropriate for determining half lives (especially of long-lived nuclei), it is inappropriate for studying variability. The TDCR procedure is based on the assumption that the beta-decay process is intrinsically isotropic, which may well be true of the basic beta-decay process, but appears not to be true of the version of the beta-decay process that is responsible for variability.

**Acknowledgments.** We express our thanks to Dr Karsten Kossert for providing us with the data analyzed in this article, for providing us with the picture shown as Figure 15, and for helpful discussion.

**Appendix. The PTB experiment**

The PTB experiment measured the *triple coincidences* of decay events as registered by three photo-multiplier tubes (PMTs). These are arranged in a horizontal plane around the source, as shown in

Figure 15. Measurements were made sequentially of the three samples (S2, S3 and S4) and also of the blank sample (S1).

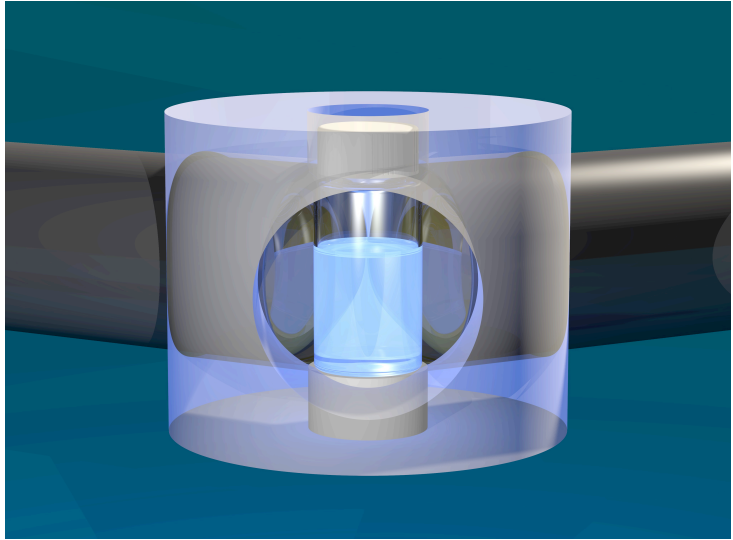

Figure 15. Simulated view of the TDCR optical system. The optical chamber consists of a cylinder with three cylindrical bores to place the photo-multiplier tubes at a minimum distance from the vial containing the radioactive specimen, and a cylindrical opening allowing the vial to be changed.

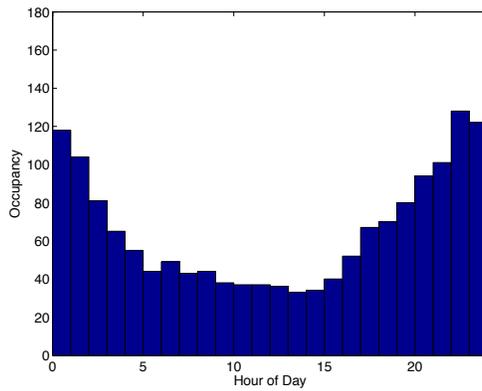

Figure 16. Histogram of hour of day of measurements of Sample 2. The vertical axis gives the number of triple coincidences as a function of hour of day.

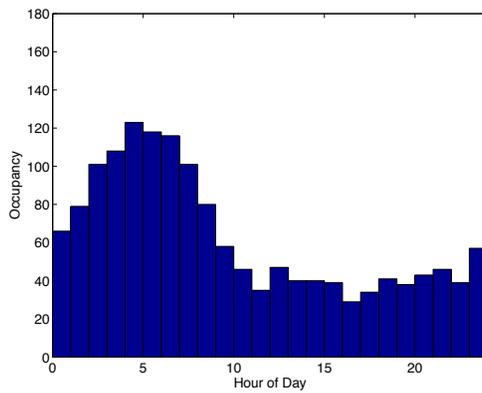

Figure 17. Histogram of hour of day of measurements of Sample 3.

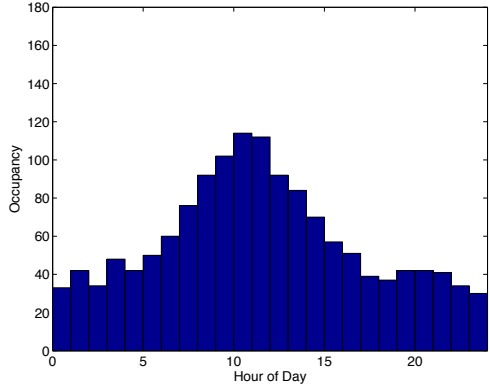

Figure 18 Histogram of hour of day of measurements of Sample 4.

Recognizing that experiments may be subject to influences that vary with time of day, we show in Figures 16, 17 and 18 hour-of-day histograms of measurements made with Samples 2, 3 and 4, respectively. We can regard these figures as plots of the "activation" of each sample as a function of time of day. We see that these plots are quite different for the three samples. Hence if the experiment is responding to any stimulus that varies with time of day, the three samples will exhibit different responses to that stimulus. The stimulus could be anything that influences the signal detected by the PMTs, for instance an environmental factor such as temperature. It follows that the differences in the power spectra for the three samples, shown in Figures 1, 2 and 3, may be due to a combination of experimental factors and the activation patterns shown in Figures 16, 17 and 18.

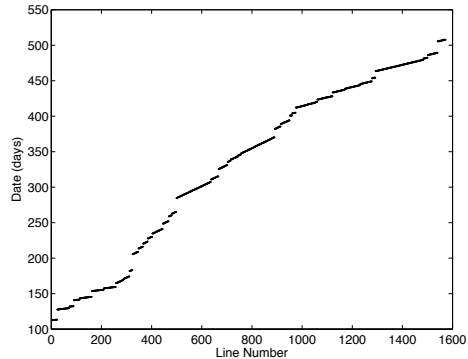

Figure 19. Plot of date of measurement in days vs line number in dataset.

Parkhomov has pointed out that *the detection of oscillations represents an experimental challenge, which requires fixed conditions of measurements with stable equipment running without interruption for many years* [31]. We show in Figure 19 a plot of the dates of measurements (in days) versus the entry numbers in the dataset (i.e. the list of discrete measurements). Had the experiment been running stably without interruption, the curve in Figure 19 would have been an unbroken line of constant slope. We see from Figure 19 that the PTB experiment does not meet the Parkhomov requirement.

References

[1] K. Kossert & O.J. Nahle, Astropart. Phys. 69, 18 (2015).
[2] E.D. Falkenberg, Apeiron, 8, No. 2, 32 (2001).
[3] G.W. Bruhn, Apeiron 9, 28 (2002).
[4] E.D. Falkenberg, Apeiron, 9, No. 2, 41 (2002).
[5] J.H. Jenkins et al., Astropart. Phys. 32, 42 (2009).
[6] E. Fischbach et al., Space Sci. Rev., 145, 285 (2009).
[7] D.E. Alburger et al., Earth Planet. Sci. Lett. 78, 168 (1986).
[8] H. Siegert et al., Appl. Radiat. Isot. 49, 1397 (1998).
[9] P.S. Cooper, Astropart. Phys. 31, 267 (2009).
[10] E.B. Norman et al., Astropart. Phys., 31, 135 (2009).
[11] T.M. Semkow et al., Phys. Lett. B, 675, 415 (2009
[12] D.E. Krause et al., Astropart. Phys. 36, 51 (2012).
[13] D. O'Keefe et al., Astrophys. Space Sci., 344, 297 (2013).
[14] J.H. Jenkins et al., Nucl. Inst. Meth. Phys. Res. A 620, 332 (2009).
[15] A. Grau Malonda et al., Appl. Radiat. Isot. 39, 1191 (1988).
[16] R. Broda et al., Appl. Radiat. Isot. 39, 159 (1988).
[17] R. Broda, Metrologia 44, 536 (2007).
[18] Parkhomov, A.G., arxiv:1004.1761 (2010).
[19] Steinitz, G., et al., Geophys. Int., 180, 651 (2010).
[20] Steinitz, G., et al., Solid Earth 1, 99 (2010).
[21] Steinitz, G., et al., Proc. Roy. Soc. A 469, 411 (2013).
[22] P.A. Sturrock et al., Phys. Rev. D 72, 11304 (2005).
[23] N. Lomb, Astrophys. Space Sci. 39, 447 (1976).
[24] J.D. Scargle, Astrophys. J. 263, 835 (1982).
[25] J.N. Bahcall & W.H. Press, Astrophys. J. 370, 730 (1991).
[26] P.A. Sturrock et al., arXiv:1510.05996 (2015).
[27] P.A. Sturrock et al., arXiv:1511.08770 (2015).
[28] Sturrock, P.A., et al., Astropart. Phys. 36, 18 (2012).
[29] Sturrock, P.A., et al., Astrophys. J. 725, 492 (2010).
[30] Sturrock, P.A., et al., Astrophys. J. 804, 47 (2015).
[31] Parkhomov, A.G., arxiv:1012.4174 (2012).